\documentclass[aps,prb,twocolumn]{revtex4-1}
\usepackage{amsmath,graphicx}
\usepackage{epstopdf}
\usepackage{color}
\usepackage{lipsum}
\usepackage{cleveref}
\emergencystretch=\maxdimen
\begin{document}
\title{Electrically tunable chiral Majorana edge modes in quantum anomalous Hall insulator-topological superconductor systems}
\author{Qing Yan$^{1,2}$}
\author{Yan-Feng Zhou$^{1,2}$}
\author{Qing-Feng Sun$^{1,2,3}$}\email{sunqf@pku.edu.cn}

\affiliation{$^{1}$International Center for Quantum Materials, School of Physics, Peking University, Beijing 100871, China}
\affiliation{$^{2}$Collaborative Innovation Center of Quantum Matter, Beijing 100871, China}
\affiliation{$^{3}$CAS Center for Excellence in Topological Quantum Computation, University of Chinese Academy of Sciences, Beijing 100190, China}

\date{\today}
\begin{abstract} 
Chiral Majorana edge modes are theoretically proposed to perform braiding operations
for the potential quantum computation.
Here, we suggest a scheme to regulate trajectories of the chiral Majorana fermion
based on a quantum anomalous Hall insulator (QAHI)-topological superconductor heterostructure.
An applied external gate voltage to the QAHI region introduces a dynamical phase
so that the outgoing Majorana fermions can be prominently tuned to different leads.
The trajectory is mechanically analyzed and the electrical manipulation is represented
by the oscillating transmission coefficients versus the gate voltage.
Through the optimization of devices, the conductance is likewise detectable
to be periodically oscillating,
which means an experimental control of chiral Majorana edge modes.
Besides, this oscillating period which is robust against disorder also provides an attainable method
of observing the energy dispersion relation of the edge mode of the QAHI.
Furthermore, the oscillating behavior of conductance
serves as smoking-gun evidence of the existence of the chiral Majorana fermion,
which could be experimentally confirmed.
\end{abstract}

\maketitle
\section{\label{sec1}Introduction}

In recent years, Majorana fermions have been a promising area of interest
in condensed matter physics, although they were first proposed to be self-conjugated elementary particles in particle physics.\cite{Alicea2012RepProgPhys,Beenakker2013annurev-conmatphys,Elliott2015RevModPhys}
Since Majorana zero modes\cite{ReadGreen2000prb,Kitaev2001}
are confirmed at the interface of topological insulators and conventional superconductors\cite{FuliangKane2008PRL,CookFranz2011PRB,SunhaohuaJiajinfeng2016PRL},
semiconductor nanowires with strong spin-orbital coupling\cite{Mourik2012Science,Zhanghao2018nature},
or magnetic atom chains\cite{Choy2011prb,Klinovaja2013PRL,NadjPerge2014Science,Pawlak2016npjqi},
they are often regarded as a probable candidate for topological quantum computing\cite{Nayak2008RevModPhys,Alicea2011naturephysics,StanescuDasSarma2018prb,Chenchuizhen2018PRBq1D}.
Besides, the chiral Majorana edge modes, as the one-dimensional (1D) homologous counterpart of Majorana zero modes,
are investigated both in theory\cite{FuliangKane2009PRL,AkhmerovBeenakker2009PRL,KTLaw2009PRL,QixiaoliangZSC2010PRBChiralTSC}
and in experiments\cite{Heqinglin2017science},
accommodated by edges of two-dimensional (2D) topological superconductors (TSCs).
Recently, the chiral Majorana edge modes are theoretically constructed
to naturally undertake the role of non-Abelian quantum gates by propagation,
hopefully to be an alternative of braiding realization\cite{Lian2018PNAS,Zhouyanfeng2019PRBnonAb}.

The quantum anomalous Hall insulator (QAHI), also known as the magnetic topological insulator,
can present chiral Dirac edge modes without an external magnetic field,
with the assistance of magnetic doping, such as Cr-$\mathrm{(Bi,Sb)_2Te_3}$
films\cite{YuruiFangzhong2010Science,Changcuizu2013science,Kouxufeng2014PhysRevLett}
or V-$\mathrm{(Bi,Sb)_2Te_3}$ films\cite{Changcuizu2015naturematerials,Sunhaohua2017SCPMA}.
These QAHIs are topologically classified by Chern number $\mathcal{C}=1$.
When the QAHI is covered by a conventional $s$-wave superconductor,
the topological superconductor emerges with $\mathcal{N}$ chiral Majorana edge modes
along each edge, denoted by Chern number $\mathcal{N}$ in the Majorana basis.
With a tiny superconducting gap, the $\mathcal{N}=2$ TSC phase is topologically
equivalent to the QAHI phase, so no backscattering occurs in
the QAHI-TSC-QAHI junction\cite{ChungSukBumZSC2011prb}.
However, as the induced superconducting gap increases, the TSC transits into
the $\mathcal{N}=1$ phase, carrying one chiral Majorana edge mode along each edge.
Here, in the QAHI-TSC-QAHI junction, backscattering comes up,
resulting in both transmission and reflection processes.
When sweeping the external magnetic field over the QAHI-TSC-QAHI junction,
He {\sl et al.}\cite{Heqinglin2017science} observe that the half-integer quantized
conductance plateau $\frac{e^2}{2h}$ occurs at the magnetization reversals,
claiming an experimental discovery of the chiral Majorana edge modes.
However, there still remain some controversies about the origin of
the half-integer conductance
plateau\cite{JiWenjieWenxiaogang2018PRL,HuangYingyi2018PRB,LianBiao2018PRB}.
Ji {\sl et al.}\cite{JiWenjieWenxiaogang2018PRL} and Huang {\sl et al.}\cite{HuangYingyi2018PRB}
independently suggest a classical interpretation of the half-integer conductance plateau
without the participation of chiral Majorana edge modes,
resulting from a good electric contact between the QAHI and the superconductor
based on a percolation model. Hence, it is earnestly expected to further demonstrate
the real existence of chiral Majorana edge modes.

When two electrons can be combined as a Cooper pair due to the attractive interaction potential,
there may come the Andreev reflection at the interface of a superconductor
and a normal conductor.\cite{Andreev1966ElectronSpectrum}
If the injected electron and the reflected hole locate at the same terminal,
this process is called the local Andreev reflection (LAR)\cite{TanakaNagaosa2009PRL}.
Or otherwise, the crossed Andreev reflection (CAR)
describes that the injected electron and reflected hole are separated between
different terminals, also known as the non-local Andreev reflection.\cite{houzhePRB2016CrossedAndreev,Deutscher2000APLCAR,Wangj2015PRBPhysQuantizedCAR}
With the induced superconducting potential, the QAHI-TSC-QAHI system can
possess the LAR and CAR transport as well as the normal tunneling transmission and reflection.
Here the non-zero Andreev reflection originates from one single chiral
Dirac edge state of the QAHI splitting into two chiral Majorana edge states at the interface of different topological regions.

In order to exploit the practical application of Majorana
fermions in realistic devices, the crucial step is to
realize the effective control and regulation of chiral Majorana states.
Considering that a Majorana fermion is a charge-neutral quasi-particle,
it should be ineffective to control and manipulate the Majorana fermions
directly by electric or magnetic fields.\cite{zhouyanfeng2018SCPMA}
In particular, the chiral Majorana fermion always flows along the edge of the TSC,
and it is difficult to control and change its flowing direction.
To date, many works have theoretically proposed various
methods.\cite{Zhangyingtao2017prb,zhouyanfeng2018prb,Chenchuizhen2018PRBJoscurrent,LichanganShenshunq2019majorana,Wangjing2018PRLmultipleNCMEM,Zengyongxin2018PRBQAHmajorana}
For example, by tuning the chemical potential or introducing a scanning
tunneling microscope tip, the phase of the chiral Majorana state can be
adjusted between zero and $\pi$.\cite{Zhangyingtao2017prb,zhouyanfeng2018prb}
In addition, Chen {\sl et al.} show that the critical Josephson current dramatically increases to a peak value in the QAHI-TSC-QAHI hybrid junctions when the TSC is in the $\mathcal{N}=1$ topological phase.\cite{Chenchuizhen2018PRBJoscurrent}
Li {\sl et al.} construct a Josephson interferometer via
a QAHI bar to exhibit a phase-dependent interference pattern.\cite{LichanganShenshunq2019majorana}
Wang {\sl et al.} argue that there exists a $\frac{2}{3} \frac{e^2}{h}$ average conductance in the QAHI-TSC-QAHI junction with a $\mathcal{N}=3$ topological phase of TSC.\cite{Wangjing2018PRLmultipleNCMEM}
Chen {\sl et al.} propose a quasi-one-dimensional QAH-TSC structure
to control Majorana zero modes so that they behave a non-Abelian time evolution.\cite{Chenchuizhen2018PRBq1D}
However, there is still a lack of effective methods to regulate
the trajectory of chiral Majorana fermions.

In this paper, we bring up an effective and easy-to-handle electrical method,
which not only can control trajectories of chiral Majorana fermions
but also causes the oscillations of the conductance
to sufficiently confirm the existence of chiral Majorana fermions.
Three QAHI-TSC-QAHI-TSC-QAHI devices (A, B, and C) are designed with the inspiration:
the topological inequivalence of QAHI and TSC
leads to Majorana states from the central QAHI being divided into two beams at the left QAHI-TSC interface.
We implement a gate voltage on the upper edge of the central QAHI,
which could add opposite dynamical phases to the propagating electron
and hole modes and continuously regulate the ejection trajectory of chiral Majorana fermions.
By the nonequilibrium Green's function technique,
the transmission and Andreev reflection coefficients are obtained and
they oscillate corresponding to the varying gate voltage
when the TSCs are in the $\mathcal{N}=1$ phase.
Then we design two improved devices to make up
for the pity of the constant conductance of Device A.
Conductances of Devices B and C exhibit the oscillation of
the conductance versus the gate voltage,
which could be measured in experiments to confirm
the existence of the chiral Majorana fermions
and exclude classical hypotheses of the half-integer conductance plateau $\frac{e^2}{2h}$.\cite{JiWenjieWenxiaogang2018PRL,HuangYingyi2018PRB,LianBiao2018PRB}
Besides, the existence of the disorders is studied and the conductance oscillation
is robust against the disorder and the superconducting gap fluctuation.
The non-zero superconducting phase difference is also discussed.
Furthermore, it is an achievable way to detect the energy dispersion relation
of chiral Dirac edge modes of QAHI based on the periodic conductance oscillation.

This paper is organized as follows. 
Section \ref{sec2} describes the model Hamiltonians of the QAHI and TSC regions,
briefly depicts the composition of devices, and concisely expounds
the transport method used in calculations.
In Secs. \ref{sec3}-\ref{sec5}, we calculate the transmission and Andreev reflection coefficients
and conductances for Devices A, B, and C, respectively,
explain how the electric gate controls chiral Majorana fermions to different leads
and show the oscillating conductance.
Section \ref{sec6} studies the effect of the disorder and the superconducting gap fluctuation
on the oscillations of the conductance.
Finally, a brief summary is presented in Sec. \ref{sec7}.

\section{\label{sec2}Model and Method}

To describe the QAHI system, we adopt a two-band effective Hamiltonian expanded
near the $\Gamma$  point\cite{QixiaoliangZSC2010PRBChiralTSC}, which is $\mathcal{H}_{\mathrm{QAHI}}=\sum_\mathbf{p}\psi_\mathbf{p}^\dagger H_{\mathrm{QAHI}}(\mathbf{p})\psi_\mathbf{p}$,
with $\psi_\mathbf{p}=(c_{\mathbf{p}\uparrow},c_{\mathbf{p}\downarrow})^T$ and,
\begin{equation}\label{HQ}
 H_{\mathrm{QAHI}}(\mathbf{p})\!=(m \!+ \! Bp^2)\sigma_z+A(p_x\sigma_x+p_y\sigma_y) \!-(\mu_\mathrm{QAHI}+V_g)\sigma_0,
\end{equation}
where $c_{\mathbf{p}\sigma}$ and $c_{\mathbf{p}\sigma}^\dag$ are, respectively,
the annihilation and creation operators with momentum $\mathbf{p}$ and spin $\sigma=\uparrow,\downarrow$.
$\sigma_{x,y,z}$ are Pauli matrices for spin and $\sigma_0$ is the 2$\times$2 identity matrix.
$A$, $B$, and $m$ are material parameters. More specifically, $A$ is related to the Fermi velocity, $B$ is the parabolic term, and $m$ denotes the mass gap.
$\mu_\mathrm{QAHI}$ describes the chemical potential,
which is set identical for the entire QAHI regions.
$V_g$ is the gate voltage, which is only non-zero within the gating QAHI region,
marked by lilac rectangles in Figs.1(a-c).
For numerical calculation, the Hamiltonian $\mathcal{H}_{\mathrm{QAHI}}$ can be further
mapped into a square lattice model in the tight-binding representation\cite{Datta1995},
\begin{align}\label{HQL}
\mathcal{H}_{\mathrm{QAHI}} = \sum_\mathbf{i}\left[\psi_\mathbf{i}^{\dag} T_0 \psi_\mathbf{i}+(\psi_\mathbf{i}^{\dag}T_x\psi_{\mathbf{i}+\delta \mathbf{x}}+\psi_\mathbf{i}^{\dag} T_y \psi_{\mathbf{i}+\delta \mathbf{y}})+\mathrm{H.c.}\right],
\end{align}
with
$T_0 =(m+4B\hbar^2/a^2)\sigma_z-(\mu_\mathrm{QAHI}+V_g)\sigma_0$,
$T_x = -(B\hbar^2/a^2)\sigma_z-(iA\hbar/2a)\sigma_{x}$ and
$T_y =-(B\hbar^2/a^2)\sigma_z-(iA\hbar/2a)\sigma_{y}$.
Here $\psi_\mathbf{i}=(c_{\mathbf{i}\uparrow},c_{\mathbf{i}\downarrow})^T$,
$c_{\mathbf{i}\sigma}$ and $c_{\mathbf{i}\sigma}^\dag$ are, respectively,
the annihilation and creation operators on site $\mathbf{i}$ with spin $\sigma$.
$a$ is the lattice length and $\delta \mathbf{x}$ ($\delta \mathbf{y}$) is
the unit cell vector along $x$ ($y$) direction.
The topological property of the Hamiltonian $\mathcal{H}_{\mathrm{\mathrm{QAHI}}}$
is determined by the sign of $m/B$.
If $m/B<0$, the QAHI state is topologically nontrivial with Chern number $\mathcal{C}=1$,
carrying one chiral edge mode along each boundary of the QAHI region.
But for $m/B>0$, the Hamiltonian $\mathcal{H}_{\mathrm{QAHI}}$ describes
a normal insulating state with Chern number $\mathcal{C}=0$.
Hereafter, we use the dimensionless parameters with $A=1$, $B=1$, $m=-0.5$, $a = 1$ and $\hbar = 1.$\cite{zhouyanfeng2018prb}

Then we place an $s$-wave superconductor on the top of the QAHI
and introduce a finite pairing potential $\Delta$ by the proximity effect.
This leads to a TSC state containing a full gap with no node,
which could be modeled in the Bogoliubov de Genns (BdG) Hamiltonian\cite{Bernevig2013TITSC},
$\mathcal{H}_{\mathrm{BdG}}=\frac{1}{2}\sum_\mathbf{p}\Psi_\mathbf{p}^\dagger H_{\mathrm{BdG}}(\mathbf{p})\Psi_\mathbf{p}$, under the basis of $\Psi_\mathbf{p}=(c_{\mathbf{p}\uparrow},c_{\mathbf{p}\downarrow}, c_{\mathbf{-p}\uparrow}^{\dag},c_{\mathbf{-p}\downarrow}^{\dag})^T$, and
\begin{equation}\label{HBDG}
H_{\mathrm{BdG}}=\left(
                      \begin{array}{cc}
                        H_{\mathrm{QAHI}}(\mathbf{p})-\mu_{\mathrm{tsc}} & i\Delta\sigma_y \\
                        -i\Delta^{\ast}\sigma_y & -H_{\mathrm{QAHI}}^{\ast}(-\mathbf{p})+\mu_{\mathrm{tsc}} \\
                      \end{array}
                    \right),
\end{equation}
where $\mu_{\mathrm{tsc}}$ describes the chemical potential of the TSC region.
In the following devices, we choose TSCs with identical material
parameters of the QAHI, and set $\Delta=0.35$.
If $m<-\sqrt{|\Delta|^2+\mu_{\mathrm{tsc}}^2}$, the TSC state lies in the $\mathcal{N}=2$ phase,
topologically equivalent to the $\mathcal{C}=1$ QAHI state.
When $m^2<|\Delta|^2+\mu_{\mathrm{tsc}}^2$, the Chern number of TSC is $\mathcal{N}=1$,
providing only one chiral Majorana mode on each boundary,
topologically different from the former case.
If $m>\sqrt{|\Delta|^2+\mu_{\mathrm{tsc}}^2}$, $H_{\mathrm{BdG}}$ describes a normal superconductor
with $\mathcal{N}=0$. Also, according to the Altland-Zirnbauer symmetry classification
scheme\cite{Altland-ZirnbauerSymmetryClassificationPRB},
the BdG Hamiltonian possesses an intrinsic particle-hole symmetry
but no time-reversal symmetry.

In this paper, we come up with three QAHI-TSC-QAHI-TSC-QAHI devices (Devices A, B, and C)
to study the propagation of chiral Majorana fermions.
Device A is a three-terminal device as shown in Fig.1(a).
From right to left, Device A is composed of the right QAHI (Lead-2), the right TSC,
the central QAHI connected to Lead-3, the left TSC, and the left QAHI (Lead-1) regions.
Device B is a two-terminal structure with two independently gating regions
along both upper and lower edges of the central QAHI region, displayed in Fig.1(b).
Device C is another three-terminal device
where the lower part of the central QAHI region is weakly coupled to Lead-2
by a quantum point contact, shown in Fig.1(c).
All of these three devices can be regarded as a central scattering region
connected to two or three leads. In our calculations, the central scattering region
contains the central QAHI region together with the left and right TSC regions,
but not including the QAHI leads, schematically shown by the dashed line box in Fig.1(a).
The leads are perfect and semi-infinite, sharing the same parameter
setting with the central QAHI region without gating.
Here we first consider that the superconducting phase difference
$\delta \varphi$ between two TSCs is zero.
In the experiment, when the two superconducting electrodes
are connected together in the external circuit,
the phase difference $\delta \varphi$ is zero
in the absence of the magnetic field.
In addition, we will investigate the non-zero phase difference $\delta \varphi$
in Sec. \ref{sec6}, and all results in this paper can well remain.

\begin{figure}
  \centering
  \includegraphics[width=1.05\linewidth]{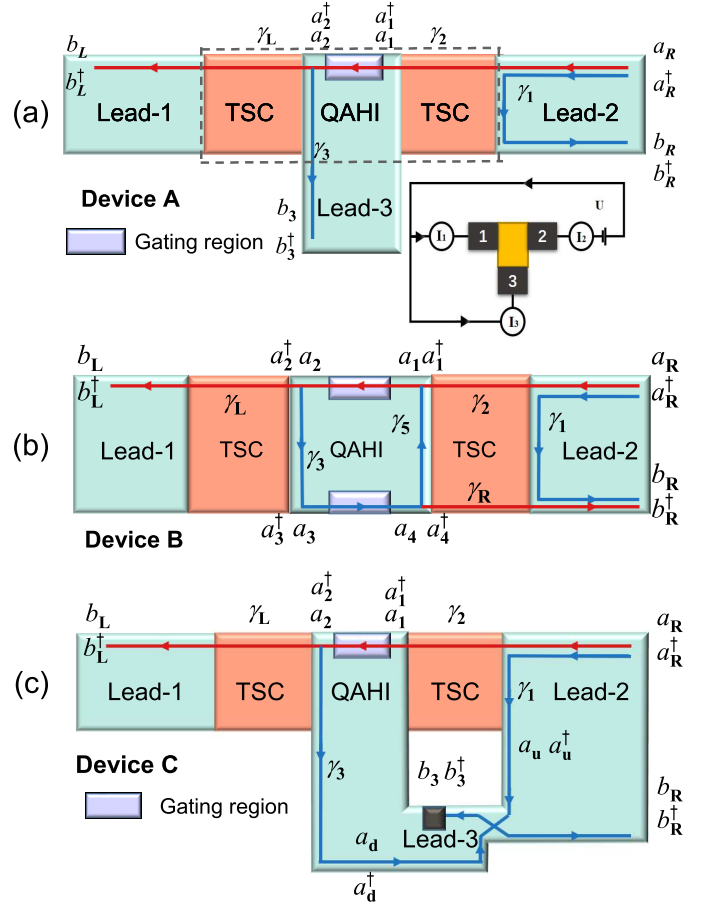}
  \caption{(color online)
(a), (b) and (c) are schematic diagrams of three QAHI-TSC-QAHI-TSC-QAHI devices
(Device A, Device B and Device C, respectively).
In the schematic diagrams,
the possible propagating routes of chiral Majorana fermions are shown,
when an electron incomes from Lead-2.
The lilac rectangle depicts the gating region.
The red and green regions are the TSC and QAHI regions, respectively.
The inset in (a) depicts the circuit layout for conductance measuring.}
\label{fig:fig1}
\end{figure}

\begin{figure}
  \centering
  \includegraphics[width=1.05\linewidth]{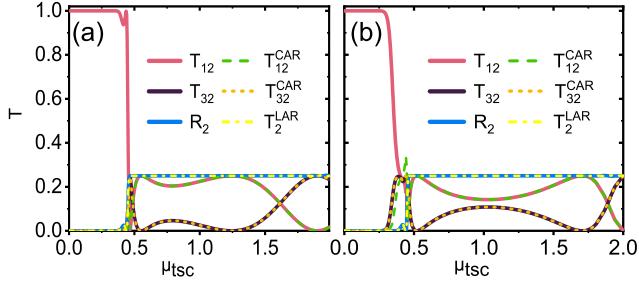}
  \caption{(color online)
The transport properties of Device A. The normal transmission coefficients $T_{12}$ and $T_{32}$,
the CAR coefficients $T^\mathrm{CAR}_{12}$ and  $T^\mathrm{CAR}_{32}$,
the normal reflection coefficient $R_2$,
and the LAR coefficient $T^\mathrm{LAR}_2$ as functions of $\mu_{\mathrm{tsc}}$,
with $\mu_\mathrm{QAHI}=0$ (a) and 0.2 (b), respectively.
The gate voltage $V_g=0$, and the length of gating region $L_v=20a$.
Here the normal transmission and reflection coefficients are plotted with solid lines,
while CAR and LAR coefficients with dashed lines,
so as to the distinction of outgoing electron or hole modes.
  }
  \label{fig:fig2}
\end{figure}

The scattering processes through two-terminal or three-terminal devices
are analyzed by the nonequilibrium Green's function technique\cite{Datta1995},
giving rise to the transmission coefficients as follows\cite{Chengshuguang2009PRL},
\begin{eqnarray}
T_{mn}(E)&=&\mathrm{Tr}[\Gamma^m_{ee}\textbf{G}^r_{ee}\Gamma^n_{ee}\textbf{G}^a_{ee}],\label{eq.T}\\
T^\mathrm{CAR}_{mn}(E)&=&\mathrm{Tr}[\Gamma^m_{ee}\textbf{ G}^r_{eh}\Gamma^n_{hh} \textbf{G}^a_{he}],\label{eq.CART}\\
T^\mathrm{LAR}_n(E)&=&\mathrm{Tr}[\Gamma^n_{ee}\textbf{G}^r_{eh}\Gamma^n_{hh}\textbf{G}^a_{he}],\label{eq.LART}
\end{eqnarray}
where $e$ and $h$ represent the electron and hole, respectively.
$E$ denotes the incident energy. $n$ and $m$ are the indices of terminals, with $n\neq m$.
$T_{mn}(E)$ and  $T^\mathrm{CAR}_{mn}(E)$ are, respectively, the normal and CAR transmission
coefficients from terminal $n$ to terminal $m$,
and $T^\mathrm{LAR}_n(E)$ is the LAR coefficient at terminal $n$.
${\bf G}^r(E)=[E-\mathcal{H}_{\mathrm{cen}}-\sum_{n} {\bf \Sigma}^r_{n}]^{-1}$
is the retarded Green's function, where $\mathcal{H}_{\mathrm{cen}}$
is the Hamiltonian of the central scattering region.
The coupling between QAHI Lead-$n$ and the center region is described
by the line-width function ${\Gamma}^{n}(E)=i[ {\bf \Sigma}_{n}^r- {\bf \Sigma}_{n}^a]$,
where the self-energy function satisfies ${\bf \Sigma}_{n}^r=[{\bf \Sigma}_{n}^{a}]^{\dagger}$.
Since there is only one edge mode injecting from each QAHI terminal,
the normal reflection coefficient at the QAHI terminal $n$ is
\begin{equation}\label{ep.R}
  R_{n}=1 - \! \sum\limits_{m(m\not= n)} \! T_{mn} \! - \! \sum\limits_{m}T^\mathrm{CAR}_{mn}-T^\mathrm{LAR}_n
\end{equation}

\section{\label{sec3}Results of Device A }

In this section, we regulate the trajectory of chiral Majorana fermions
by a designed Device A, based on the QAHI-TSC hybrid systems.
Initially, we construct a QAHI-TSC-QAHI-TSC-QAHI device shown in Fig.1(a). The central scattering region, TSC-QAHI-TSC, is connected to three semi-infinite QAHI leads with widths $100a$, $100a$, and $60a$, respectively, labeled by Lead-1, Lead-2, and Lead-3.
The length of two identical TSCs is $80a$.
On the top of the centeral QAHI, an external gate voltage, $V_g$,
is applied along the upper edge to tune the chemical potential.
The gating region is colored lilac, of which the length, $L_v$,
equals $20a$ unless specified otherwise.
The width of the gating region, $W_v$, values $20a$,
which is much longer than the broadening width of the chiral Dirac edge modes,
ensuring that the Dirac edge carriers travel smoothly through this gating region.
The energy of an incident electron from Lead-2 is fixed at $E=0$.

We discuss the transport process with the beginning of one electron mode from Lead-2,
amount to two Majorana fermions.
Trajectories of the injecting Majorana fermions are analyzed
based on all the transmission coefficients and Andreev reflection coefficients
through the three-terminal Device A,
specifically labeled by $T_{12}$,  $T_{32}$, $R_2$, $T^\mathrm{CAR}_{12}$,
$T^\mathrm{CAR}_{32}$, and $T^\mathrm{LAR}_2$.

In order to illustrate the physical picture of the injecting Majorana fermions
from Lead-2, we first study the transport without the gate voltage.
Figure 2(a) shows the transmission coefficients and Andreev reflection coefficients versus
the chemical potential of TSC, $\mu_{\mathrm{tsc}}$.
From Fig.2(a), when $\mu_{\mathrm{tsc}}$ is less than a critical value $\mu_{\mathrm{tsc}}^c$,
approximately 0.5 in our parameters,
only the normal transmission coefficient from Lead-2 to Lead-1,
$T_{12}$, is not zero with $T_{12} =1$, but $T_{32} =R_2 = T^\mathrm{CAR}_{12}
=T^\mathrm{CAR}_{32} = T^\mathrm{LAR}_2 =0$.
Because the TSC locates in the $\mathcal{N}=2$ TSC phase at $\mu_{\mathrm{tsc}}<\mu_{\mathrm{tsc}}^c$,
which is topologically equivalent to the $\mathcal{C}=1$ QAHI.
In this case, the carrier perfectly propagates via the chiral edge modes,
counterclockwise from Lead-2 to Lead-1, from Lead-1 to Lead-3,
and from Lead-3 to Lead-2.
Given an incident electron from Lead-2, it would perfectly propagate
to Lead-1 through the central TSC-QAHI-TSC region, as is the explanation for $T_{12}=1$,
and $T_{32} =R_2 = T^\mathrm{CAR}_{12} =T^\mathrm{CAR}_{32} = T^\mathrm{LAR}_2 =0$ in Fig.3(a).
On the other hand, once $\mu_{\mathrm{tsc}}$ exceeds the critical value $\mu_{\mathrm{tsc}}^c$,
the TSC will jump into the $\mathcal{N}=1$ phase,
and all the transmission coefficients and Andreev reflection coefficients appear
with $T_{12}=T^\mathrm{CAR}_{12}$, $T_{32}=T^\mathrm{CAR}_{32}$,
and $R_2=T^\mathrm{LAR}_2$ as shown in Fig.2(a).
Here $R_2=T^\mathrm{LAR}_2 =\frac{1}{4}$ is always true regardless of the chemical potential $\mu_{\mathrm{tsc}}$,
but $T_{12}$, $T^\mathrm{CAR}_{12}$, $T_{32}$, and $T^\mathrm{CAR}_{32}$ depend on $\mu_{\mathrm{tsc}}$.
This means that one of the two injecting Majorana fermions from Lead-2 is totally reflected back to Lead-2
and the other is transmitted to Lead-1 or Lead-3.
Besides, when we adjust the chemical potential of QAHI, $\mu_\mathrm{QAHI}$, the results can well remain.
For example, Fig.2(b) shows the six transmission coefficients at $\mu_\mathrm{QAHI}=0.2$.
Here $T_{12} =1$ and $T_{32} =R_2 = T^\mathrm{CAR}_{12}
=T^\mathrm{CAR}_{32} = T^\mathrm{LAR}_2 =0$ at $\mu_{\mathrm{tsc}}<\mu_{\mathrm{tsc}}^c$, and all the six transmission coefficients appear at $\mu_{\mathrm{tsc}}>\mu_{\mathrm{tsc}}^c$, which is very similar to that in Fig.2(a).
Below, we concentrate on the $\mathcal{N}=1$ TSC phase to see the novel quantum oscillation.

\begin{figure}
  \centering
  \includegraphics[width=1\linewidth]{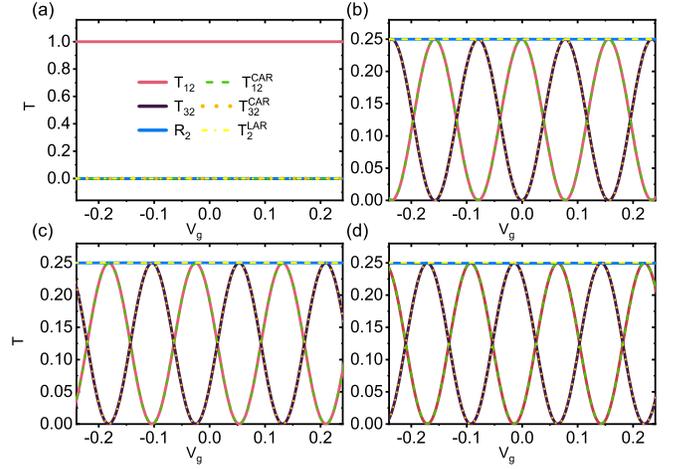}
\caption{(color online)
$T_{12}$, $T_{32}$, $R_2$, $T^\mathrm{CAR}_{12}$,
$T^\mathrm{CAR}_{32}$, and $T^\mathrm{LAR}_2$ versus the gate voltage $V_g$
for the different chemical potentials of TSC,
$\mu_{\mathrm{tsc}}=0.2,\ 0.55,\ 1,\ 2$ for (a), (b), (c), and (d) respectively.
All the unmentioned parameters are the same as Fig.2(a).}
  \label{fig:fig3}
\end{figure}

Figure 3 shows the normal transmission coefficients and Andreev reflection coefficients
versus the gate voltage $V_g$ at several specific chemical potentials $\mu_{\mathrm{tsc}}$.
When $\mu_{\mathrm{tsc}}=0.2<\mu_{\mathrm{tsc}}^c$, the normal transmission coefficient $T_{12}=1$ exactly
and $T_{32} =R_2 = T^\mathrm{CAR}_{12} =T^\mathrm{CAR}_{32} = T^\mathrm{LAR}_2 =0$ [see Fig.3(a)],
because the TSC locates in the $\mathcal{N}=2$ TSC phase.
On the other hand, when the TSC is in the $\mathcal{N}=1$ TSC phase
with $\mu_{\mathrm{tsc}}>\mu_{\mathrm{tsc}}^c$ in Figs.3(b-d),
the normal reflection and the LAR coefficients $R_2=T^\mathrm{LAR}_2 =\frac{1}{4}$.
In particular, the normal transmission coefficients ($T_{12}$ and $T_{32}$)
and the CAR coefficients ($T^\mathrm{CAR}_{12}$ and $T^\mathrm{CAR}_{32}$)
oscillate with the gate voltage $V_g$.
The oscillations always maintain themselves as long as $\mu_{\mathrm{tsc}}>\mu_{\mathrm{tsc}}^c$ [see Figs.3(b-d)].
The amplitude of the oscillation is $1/4$.
When $T_{12}$ and $T^\mathrm{CAR}_{12}$ oscillate to the maximum $1/4$,
$T_{32}$ and $T^\mathrm{CAR}_{32}$ oscillate to the minimum $0$, and vice versa.
In addition, $T_{12}=T^\mathrm{CAR}_{12}$, $T_{32}=T^\mathrm{CAR}_{32}$,
and $T^\mathrm{CAR}_{12} +T^\mathrm{CAR}_{32} =1/4$.

Let us analyze the trajectory of chiral Majorana fermions,
discuss the control of their propagating route,
and explain results in Fig.3 when TSC is in the $\mathcal{N}=1$ phase.
To begin with, a Dirac electron injects from Lead-2,
which is tantamount to two Majorana fermions,
$a_{R}=\frac{\sqrt{2}}{2}(\gamma_1+i\gamma_2)$.
At the boundary of Lead-2 and the right TSC, one of the Majorana fermions
$\gamma_1$ is totally reflected, travels down and backscatters to Lead-2.
Thus, the outgoing electron and hole modes appear equiprobably, $\gamma_1=\frac{\sqrt{2}}{2}(b_{R}+b^{\dag}_{R})$,
and we can obtain the normal reflection and LAR coefficients,
\begin{equation}
  R_2=T^\mathrm{LAR}_2=1/4.
\end{equation}
On the other hand, the other Majorana fermion $\gamma_2$ passes through the right TSC
and reaches the central QAHI region as mixing of electron and hole,
$\gamma_2=\frac{\sqrt{2}}{2i}(a_{1}-a^{\dag}_{1})$, as shown in Fig.1(a).
Then passing the central gating region,
the electron mode $a_{1}$ can acquire a dynamical phase $\phi$ and become $a_{2}$.
At the same time, the hole mode $a^{\dag}_{1}$ gains an opposite phase $-\phi$
and thus becomes $a^{\dag}_{2}$,
\begin{equation}\label{Aa2-a1}
  a_{2}=e^{i\phi}a_{1},\  a^{\dag}_{2}=e^{-i\phi}a^{\dag}_{1},
\end{equation}
where $\phi$ is the dynamical phase controlled by the gate voltage $V_g$.
Then, again at the QAHI-left TSC boundary,
splitting happens: one Majorana fermion $\gamma_L$ is transmitted to Lead-1, $\gamma_L=\frac{\sqrt{2}}{2i}(a_{2}-a^{\dag}_{2})$,
while the other Majorana fermion $\gamma_3$ is reflected to Lead-3, $\gamma_3=\frac{\sqrt{2}}{2}(a_{2}+a^{\dag}_{2})$.
The outgoing Majorana modes have the relations to the original Majorana mode $\gamma_2$ as,
\begin{equation}\label{Eq:gamma3_gammL}
  \gamma_L=\gamma_2\sin\phi,\  \gamma_3=\gamma_2\cos\phi.
\end{equation}
When $\phi=0$, $\gamma_3=\gamma_2$, the Majorana state $\gamma_2$ totally ejects into Lead-3,
but when $\phi=\frac{\pi}{2}$, $\gamma_L=\gamma_2$,
which means $\gamma_2$ entirely departs via Lead-1.
With other values of the phase $\phi$, the incoming Majorana fermion $\gamma_2$
is controlled to leave partially through Lead-1 and Lead-3.
So we can well control the propagating route of the chiral Majorana fermion
by tuning the phase $\phi$.
Eventually, the Majorana fermions $\gamma_L$ and $\gamma_3$ respectively eject into Lead-2 and Lead-3, leading that the normal transmission coefficients and CAR coefficients are,
\begin{eqnarray}
  T_{12}=T^\mathrm{CAR}_{12}=(1/4){\sin}^2(\phi+\phi_0),\\
  T_{32}=T^\mathrm{CAR}_{32}=(1/4){\cos}^2(\phi+\phi_0),
\end{eqnarray}
and $T_{12}+T_{32}=T^\mathrm{CAR}_{12} +T^\mathrm{CAR}_{32} =1/4$, where $\phi_0$ denotes the initial phase.
The sinusoidal oscillations of the normal transmission coefficients
and CAR coefficients are well consistent with the numerical curves in Figs.3(b-d).
Notice that here the normal transmission coefficient $T_{12}$ ($T_{32}$) is always
equal to the CAR coefficient $T^\mathrm{CAR}_{12}$ ($T^\mathrm{CAR}_{32}$),
because the outgoing Majorana fermion $\gamma_L$ ($\gamma_3$) to Lead-1 (Lead-3)
has the same components of electron and hole.
It is worth to mention that if the right TSC is removed,
the single QAHI-TSC-QAHI device can no longer control the chiral Majorana fermion
to Lead-1 and Lead-3 by tuning the gate voltage, stemming from the coexistence of
$\gamma_1$ and $\gamma_2$ in the right QAHI.
Remarkably, the two-TSC structure with $\mathcal{N}=1$ in Device A is essential
to control chiral Majorana fermions into different terminals via electrical gating.

Then let us explain how the gate voltage modulates trajectories of Majorana modes continuously.
When a gate voltage $V_g$ is applied to the upper edge of
the central QAHI region, it directly adjusts the Fermi level of QAHI,
and thus the momenta $k$ of electrons and holes with energy $E=0$ are changed.
By varying the gate voltage, the obtained dynamical phase $\phi$,
explicit to be ${kL_v}$ with $L_v$ the length of the gating region,
can tune the probabilities that the incoming Majorana mode $\gamma_2$ chooses to
go to Lead-1 or Lead-3, as Eq.(\ref{Eq:gamma3_gammL}).
Under a fixed length $L_v$, the normal tunneling coefficients $T_{12}$ and $T_{32}$
sinusoidally oscillate in the same period,
in response to a continuously varying $V_g$, as depicted in Figs.3(b-d).

If the TSC locates in the $\mathcal{N}=1$ TSC phase with $\mu_{\mathrm{tsc}}>\mu_{\mathrm{tsc}}^c$,
the chemical potential $\mu_{\mathrm{tsc}}$ of TSC has no influence on the period and
amplitude of the transmission oscillation.
From Figs.3(b-d), we can see that the oscillation amplitudes
are always $1/4$ regardless of $\mu_{\mathrm{tsc}}$.
Here the chemical potential $\mu_{\mathrm{tsc}}$ merely changes
the initial phase $\phi_0$ of the oscillation of transmission coefficients
versus $V_g$, by comparing Figs.3(b), (c), and (d).
This phase shift originates from the fact  that a Dirac wave function requires
a matching condition with the Majorana wave functions at the left TSC-central QAHI boundary
due to the transverse broadening of wave functions.

\begin{figure}
\centering
\includegraphics[width=1\linewidth]{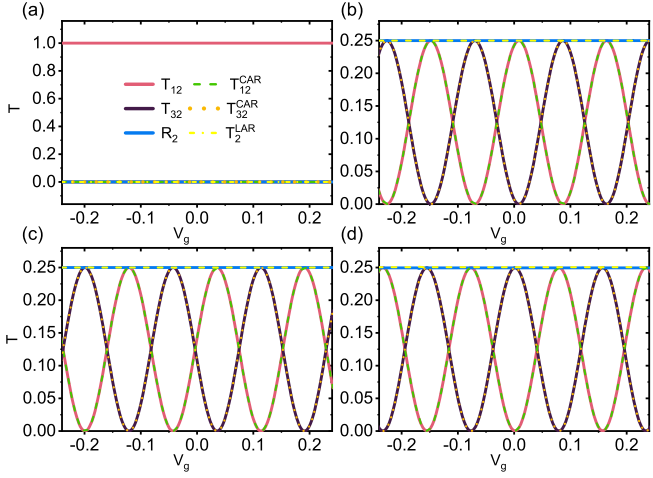}
\caption{(color online)
$T_{12}$,  $T_{32}$, $R_2$, $T^\mathrm{CAR}_{12}$,
$T^\mathrm{CAR}_{32}$, and $T^\mathrm{LAR}_2$ versus the gate voltage $V_g$
for the different chemical potentials of TSC,
$\mu_{\mathrm{tsc}}=0.2,\ 0.55,\  1, \ 2$ for (a), (b), (c), and (d) respectively.
All the unmentioned parameters are the same as Fig.2(b).}
\label{fig:fig4}
\end{figure}

The chemical potential $\mu_\mathrm{QAHI}$ of the QAHI region does not affect
on the oscillating behavior of the transmission coefficients,
but just arouses a horizonal shift compared to the initial curves.
By increasing $\mu_\mathrm{QAHI}$ from 0 to 0.2,
we plot transmission coefficients in Fig.4 as a comparison for Fig.3.
In Fig.4(a) $\mu_{\mathrm{tsc}}=0.2<\mu_{\mathrm{tsc}}^c$, the TSC is at the $\mathcal{N}=2$ TSC phase,
here only $T_{12}=1$ and other transmission and Andreev reflection coefficients are zero, as is in complete agreement with Fig.3(a).
In this case, two Majorana fermions are totally transmitted from Lead-2
into Lead-1 without reflection.
As the chemical potential $\mu_{\mathrm{tsc}}$ increases,
different topological properties between the QAHI and $\mathcal{N}=1$ TSC
appear so that Device A can regulate the Majorana mode via electrical gating as before.
Now the transmission coefficients ($T_{12}$ and $T_{32}$)
and the CAR coefficients ($T^\mathrm{CAR}_{12}$ and $T^\mathrm{CAR}_{32}$) 
oscillate with the gate voltage $V_g$, which is consistent with that in Figs.3(b-d).
However, due to the increase of the chemical potential $\mu_\mathrm{QAHI}$,
the momenta $k$ of zero-energy electrons and holes through the gating QAHI region
are no longer zero when $V_g=0$.
It behaves like an additional phase $e^{i\phi_\mathrm{QAHI}}$ to $a_{1}$,
and $e^{-i\phi_\mathrm{QAHI}}$ to $a^\dagger_{1}$.
So the transmission and Andreev reflection coefficients are,
\begin{eqnarray}
    R&=&T^\mathrm{LAR}=1/4,\\
    T_{12}&=&T^\mathrm{CAR}_{12}=(1/4){\sin}^2(\phi+\phi_0+\phi_\mathrm{QAHI}),\\
    T_{32}&=&T^\mathrm{CAR}_{32}=(1/4){\cos}^2(\phi+\phi_0+\phi_\mathrm{QAHI}).
  \end{eqnarray}
Or rather to say, the non-zero $\mu_\mathrm{QAHI}$ can be regarded
as an initial gating voltage, and thus it does not influence the period and
amplitude of any transmission oscillation.
The phase shift can be clearly seen by comparing Figs.4(b-d) with Figs.3(b-d), respectively.

\begin{figure}
\centering
\includegraphics[width=0.9\linewidth]{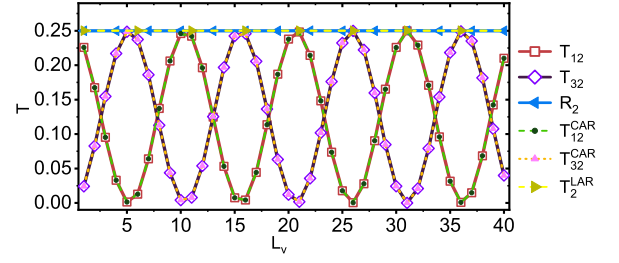}
\caption{(color online)
$T_{12}$, $T_{32}$, $R_2$, $T^\mathrm{CAR}_{12}$, $T^\mathrm{CAR}_{32}$,
and $T^\mathrm{LAR}_2$ as functions of the length of gating region $L_v$,
where $\mu_\mathrm{QAHI}=0$, $\mu_{\mathrm{tsc}}=0.55$, and $V_g=0$.
All the unmentioned parameters are the same as Fig.2(a).}
\label{fig:fig5}
\end{figure}

In Fig.5, we plot the transmission coefficients
as a function of the length $L_v$ of the gating region in the central QAHI,
and find that $R_2$ and $T^\mathrm{LAR}_2$ keep a constant $1/4$
and the transmission coefficients behave periodically oscillating versus the length $L_v$ because of the dynamical phase $\phi=k L_v$.
In addition, we change the length of the gating region,
and find that oscillations of transmission coefficients
in Figs.3 and 4 always exist. It is the length $L_v$ that determines the period of the oscillation.
Since the dynamical phase has an explicit form, $e^{ikL_v}$,
the identical oscillating period in Figs.3 and 4 is relevant to $\frac{2\pi}{L_v}$.
The longer $L_v$ is, the faster the oscillation becomes versus the gate voltage $V_g$.

\begin{figure}
\centering
\includegraphics[width=1\linewidth]{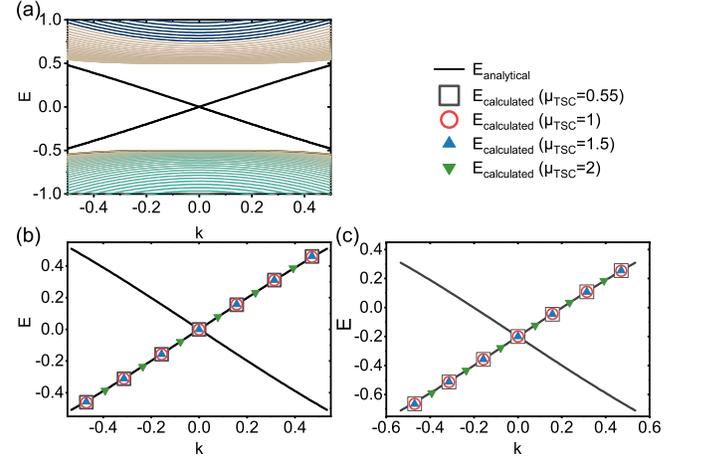}
\caption{(color online)
(a) Band structure of the QAHI nanoribbon with the width of $100a$ and $\mu_\mathrm{QAHI}=0$.
Black lines are bolded for the edge states of QAHI.
(b) and (c) are the energy dispersions obtained from the transmission coefficients in Figs.3(b-d)
and Figs.4(b-d), with $\mu_\mathrm{QAHI}=0$ and $\mu_\mathrm{QAHI}=0.2$, respectively.
The black solid lines in (b) and (c) are the analytical edge states directly from
(a) for comparison.
All the unmentioned parameters are the same as Fig.2.
}
\label{fig:fig6}
\end{figure}

Furthermore, the energy dispersion of the chiral Dirac edge states of
the central QAHI could be obtained according to the periodic
oscillation of transmission coefficients versus the gate voltage.
From the Hamiltonian of QAHI with $\mu_\mathrm{QAHI}=0$ in Eq.(\ref{HQL}),
we can directly calculate the energy dispersion
by considering a QAHI nanoribbon, as shown in Fig.6(a).
Two chiral Dirac edge states, bolded in Fig.6(a), traverse across
the bulk band gap, exactly at $E=0$ and $k=0$.
From the periodic oscillation of transmission coefficients in Fig.3,
we can obtain the energy dispersion of edge states.
While the chemical potential $\mu_\mathrm{QAHI}=0$ and the gate voltage $V_g=0$,
the momentum $k$ is set to zero.
With a given gating length $L_v$, after $n$ periodic oscillations
(with the integer $n=0,\ \pm1,\ \pm2$, ...), the momentum is $k_n=n\pi/L_v$
originating from the period of $T_{12}=(1/4){\sin}^2(kL_v)$,
and the corresponding energy $E_n$ can be obtained
from the periodic oscillation curves in Fig.3, with $E_n=V_{gn}$
where $V_{gn}$ is the value of the gate voltage after $n$ periodic oscillations.
Based on the data $(E_n,\ k_n)$, we can plot a series of discrete points in Fig.6(b).
For comparison, the analytical energy dispersion of edge states from Fig.6(a)
is also displayed in Fig.6(b).
The dispersion relation obtained from transmission coefficients
conforms perfectly to the analytical black curve.
This provides an effective method to measure the energy dispersion relation
of the chiral Dirac edge state experimentally.
Notably, this measuring method is free from different
TSC's chemical potential $\mu_{\mathrm{tsc}}$,
shown in Fig.6(b) where all the markers obtained from
the different $\mu_{\mathrm{tsc}}$ locate on the same analytical curve.
Considering the initial chemical potential $\mu_\mathrm{QAHI}$ of QAHI,
the varying $\mu_\mathrm{QAHI}$ just moves down the Fermi level
and changes the momenta of Dirac fermions in the gating region
but has no effects on the structure of the dispersion relation,
so the calculated $E-k$ relation of $\mu_\mathrm{QAHI}=0.2$
could also be obtained only with a shift along $E$ axis in Fig.6(c),
and the discrete data points are obtained from Figs.4(b-d).

Finally, we calculate the conductance of Device A.
Using the multi-probe Landauer-B\"{u}ttiker formula,
the current in Lead-$n$ at the small bias limit can be calculated,\cite{Sunqf1999PRB,sunqingfeng2009JPCM}
\begin{eqnarray} \label{Eq:LBformula}
   I_n &= & \frac{e^2}{h}\left[ \sum_{m(m\not= n)}(V_n-V_m)T_{mn}
   \right. \nonumber \\
     & & \left. + 2V_nT^\mathrm{LAR}_{n}  + \sum_{m(m\not= n)}(V_n+V_m)T^\mathrm{CAR}_{mn}
     \right] \nonumber\\
     &=&\! \frac{e^2}{h}\!\left[\! V_n(1 \! +\! T_n^\mathrm{L\!A\!R}\!-\! R_n)
     \!+\!\sum_{m(\!m\!\not=\! n\!)}\! V_m(T_{\!m\!n}^\mathrm{C\!A\!R}\!-\! T_{\!m\!n}) \!\right]\!,
\end{eqnarray}
where $V_n$ is the voltage of Lead-$n$,
and voltages of superconductors are set to zero.
Based on the circuit in the inset of Fig.1(a)
to measure the three-terminal conductance of Device A,
the linear conductances $G_n$ ($G_n\equiv I_n/V_n$ with $n=1,2,3$) at all three leads
are constant with $G_1=G_2=G_3=\frac{e^2}{h}$,
and $G_n$ does not oscillate with the gate voltage, when the TSC is in the $\mathcal{N}=1$ TSC phase.
In fact, regardless of the connection of the external circuit,
the conductance $G_n$ of Device A is always constant, although the normal transmission
and Andreev reflection coefficients oscillate with the gate voltage.
In the $\mathcal{N}=1$ TSC phase,
the two injecting chiral Majorana fermions from Lead-$n$ respectively
transmit into Lead-2, Lead-1, and Lead-3,
so that the outgoing electron and hole in all three leads are equiprobable,
leading to the relations of the transmission coefficients and Andreev reflection coefficients:
\begin{eqnarray}
  T_{mn}=T^\mathrm{CAR}_{mn}, \hspace{2mm} & &
  R_n=T^\mathrm{LAR}_{n},
\end{eqnarray}
as shown in Figs.3(b-d) and Figs.4(b-d).
Thus, $I_n =(e^2/h)V_n$ from Eq.(\ref{Eq:LBformula}),
and Device A gives no contribution to the observable
oscillation of the conductance versus the gate voltage.
Motivated by the desire for an experimentally observable oscillation of physical quantity,
we proceed with the analysis of Devices B and C in the following sections.

\section{\label{sec4}Results of Device B }

\begin{figure}
\centering
\includegraphics[width=1.05\linewidth]{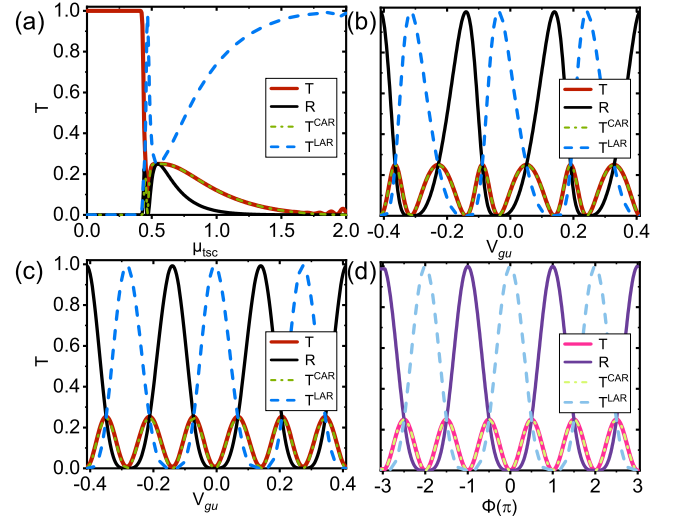}
\caption{(color online)
The transport properties of Device B.
(a) The normal tunneling coefficients $T$, the CAR coefficients $T^\mathrm{CAR}$,
the normal reflection coefficient $R$, and the LAR coefficient $T^\mathrm{LAR}$
as functions of the chemical potential $\mu_{\mathrm{tsc}}$ of the TSC,
with $\mu_\mathrm{QAHI}=0$, $L_v=20a$, $V_{gu}=V_{gl}=0$.
(b) and (c) are the transmission coefficients versus the upper gating voltage $V_{gu}$ with $\mu_{\mathrm{tsc}}=1.5$ and $1.8$, respectively.
All the unmentioned parameters are the same as in (a).
(d) The analytical transmission coefficients versus the dynamical phase $\phi$
derive from the Eqs.(\ref{B_transT}-\ref{B_transLAR}),
setting $\phi_u=\phi+\frac{\pi}{2}$ and $\phi_l=\frac{\pi}{2}$,
fitting with curves in Fig.7(c)}
\label{fig:fig7}
\end{figure}

For the purpose of observing the manipulation of
the chiral Majorana fermions in the experiments,
in this section, we design a two-terminal
QAHI-TSC-QAHI-TSC-QAHI Device B as shown in Fig.1(b).
Below, we study the transport properties of chiral Majorana fermions
both analytically and numerically, and then successfully obtain
the observable conductance oscillation as well as
the successive energy dispersion of the chiral Dirac edge modes.
In Device B, an additional lower gate in the central QAHI region is added,
and it plays the same role as the upper gate [see Fig.1(b)].
More importantly, it is a cyclic trajectory of the
chiral Majorana edge mode locating in the central QAHI region that causes
the conductance oscillation.

Similarly, for Device B, the normal transmission coefficient $T$,
the normal reflection coefficient $R$
and the CAR and LAR coefficients ($T^\mathrm{CAR}$ and $T^\mathrm{LAR}$)
can be calculated from Eqs.(\ref{eq.T}-\ref{ep.R}).
Fig.7(a) shows $T$, $R$, $T^\mathrm{CAR}$, and $T^\mathrm{LAR}$ as functions of
the chemical potential $\mu_{\mathrm{tsc}}$ of the TSC region.
When $\mu_{\mathrm{tsc}}<\mu_{\mathrm{tsc}}^c$, the TSC is in the $\mathcal{N}=2$ TSC phase,
with the normal transmission coefficient $T=1$ and $T^\mathrm{CAR}=R=T^\mathrm{LAR}=0$. These stable coefficients occur
because the injecting Dirac electron from the right QAHI,
which is equivalent to two injecting chiral Majorana fermions,
passes through the center TSC-QAHI-TSC region and directly ejects from the left QAHI.
However, when the chemical potential $\mu_{\mathrm{tsc}}>\mu_{\mathrm{tsc}}^c$,
the TSC is in the $\mathcal{N}=1$ TSC phase. Thus, the reflection of
chiral Majorana fermions occurs at the interface of the QAHI and TSC,
resulting in the transport process where transmission coefficients $T$, $R$,
$T^\mathrm{CAR}$, and $T^\mathrm{LAR}$ are usually non-zero.
From hereon, we focus on the $\mathcal{N}=1$ TSC phase with $\mu_{\mathrm{tsc}}>\mu_{\mathrm{tsc}}^c$.
Figs.7(b and c) show the transmission coefficients versus the upper gate
voltage $V_{gu}$ for the fixed chemical potential $\mu_{\mathrm{tsc}}=1.5$ and $1.8$, respectively.
All of the four transmission coefficients ($T$, $R$, $T^\mathrm{CAR}$, and $T^\mathrm{LAR}$)
oscillate with the gate voltage $V_{gu}$.
The oscillation amplitude of the normal reflection and LAR coefficients is about 1
but that of the normal transmission and CAR coefficients is about 1/4.
The oscillation period of $R$ and $T^\mathrm{LAR}$ is twice longer than that of $T$ and $T^\mathrm{CAR}$.
In addition, $T=T^\mathrm{CAR}$ always keeps, but $R$ is not equal to $T^\mathrm{LAR}$ as usual.
The oscillation of transmission coefficients versus $V_{gu}$
can well remain regardless of the chemical potentials
($\mu_{\mathrm{tsc}}$, $\mu_\mathrm{QAHI}$) and the lower gate voltage $V_{gl}$,
as long as the TSC is in the $\mathcal{N}=1$ TSC phase.
Different from Device A, the chemical potential $\mu_{\mathrm{tsc}}$ does not only shift the initial phase of transmission curves
but also obviously changes the shape of curves [see Figs.7(b and c)].

Let us analyze the propagating route of chiral Majorana fermions
and give the analytical expressions of the normal transmission and
Andreev reflection coefficients of Device B in the $\mathcal{N}=1$ TSC phase.
Considering that an electron $a_R$ from Lead-2 propagates along the upper side of QAHI,
also regarded as two chiral Majorana fermions $\gamma_1$ and $\gamma_2$
with $\gamma_1=\frac{\sqrt{2}}{2}(a_R+a_R^{\dagger})$
and $\gamma_2=\frac{\sqrt{2}}{2i}(a_R-a_R^{\dagger})$.
When they reach the right QAHI-TSC interface, $\gamma_2$ passes through the right TSC
while $\gamma_1$ is reflected back to the right QAHI, i.e., Lead-2 [see Fig.1(b)].
Then $\gamma_2$ leaves along the edge of the right TSC and enters
into the central QAHI with upper and lower gates.
To avoid confusion, let us label the four electron modes at
the vertices of the central QAHI region as $a_1$, $a_2$, $a_3$, and $a_4$,
from the upper right to the lower right in the counterclockwise direction,
and similarly the hole modes as $a^{\dag}_1$, $a^{\dag}_2$, $a^{\dag}_3$, and $a^{\dag}_4$.
The initially incoming electron $a_1$ can travel along
the upper gating QAHI edge and get a dynamical phase $\phi_u$,
while the hole $a^{\dag}_1$ acquires the phase $-\phi_u$, that is,
\begin{equation}\label{Ba2-a1}
  a_{2}=e^{i\phi_u}a_{1},\  a^{\dag}_2=e^{-i\phi_u}a^{\dag}_1.
\end{equation}
The combination of $a_2$ and $a^{\dag}_2$ splits two chiral Majorana fermions
$\gamma_L=\frac{\sqrt{2}}{2i}(a_{2}-a^{\dag}_{2})$ and
$\gamma_3=\frac{\sqrt{2}}{2}(a_{2}+a^{\dag}_{2})$.
$\gamma_L$ directly goes through the left TSC into Lead-1 with $b_L=\frac{i\sqrt{2}}{2}\gamma_L$
and $b^{\dag}_L=\frac{-i\sqrt{2}}{2}\gamma_L$,
while the other Majorana fermion $\gamma_3$ moves down counterclockwise.
Also, at the lower left vertex of the central QAHI,
$\gamma_3$ could be seen as the combination of the electron mode $a_3$
and the hole mode $a^{\dag}_3$ with
$a_3=\frac{\sqrt{2}}{2}\gamma_3$ and $a^{\dag}_3=\frac{\sqrt{2}}{2}\gamma_3$,
which acquire opposite phases $\phi_l$ and $-\phi_l$, respectively,
when passing along the lower QAHI edge,
\begin{equation}\label{a4-a3}
  a_{4}=e^{i\phi_l}a_{3},\  a^{\dag}_4=e^{-i\phi_l}a^{\dag}_3.
\end{equation}
Afterward, the gated modes $a_4$ and $a^{\dag}_4$ can again
be seen as two Majorana modes, $a_{4}=\frac{\sqrt{2}}{2}(\gamma_5+i\gamma_R)$.
$\gamma_5$ is reflected back to $a_1$,
while $\gamma_R$ ejects out and eventually combines with $\gamma_1$
to produce the outgoing modes of Lead-2
as $b_R=\frac{\sqrt{2}}{2}(\gamma_1+i\gamma_R)$ and $b^{\dag}_R=\frac{\sqrt{2}}{2}(\gamma_1-i\gamma_R)$.
Notice that the chiral Dirac electron $a_1$ and the hole $a_1^{\dagger}$
at the upper right vertex of the central QAHI originate from the
Majorana fermions $\gamma_2$ and $\gamma_5$,
\begin{eqnarray}\label{a1_gamma5}
    a_1=(\sqrt{2}/2)(\gamma_5+i\gamma_2),\ a^{\dag}_1=(\sqrt{2}/2)(\gamma_5-i\gamma_2).
\end{eqnarray}
Then we combine equations all above [from Eq.(\ref{Ba2-a1}) to Eq.(\ref{a1_gamma5})],
adopt the scattering matrix method, and express
the outgoing modes ($b_L, b_L^{\dagger}, b_R, b_R^{\dagger}$)  by
the incoming modes ($a_R, a_R^{\dagger}$) as follows:
\begin{equation}\label{Eq:IItrans1}
\left(
\begin{array}{cc}
b_L \\
b^{\dag}_L\\
 \end{array} \right)=\frac{1}{2}X
 \left(
 \begin{array}{cc}
 1 & -1\\
 -1 & 1\\
  \end{array} \right)
\left(
\begin{array}{cc}
a_R \\
a^{\dag}_R\\
 \end{array} \right),
\end{equation}
\begin{equation}\label{Eq:IItrans2}
\left(
\begin{array}{cc}
b_R \\
b^{\dag}_R\\
 \end{array} \right)=\frac{1}{2}
 \left(
 \begin{array}{cc}
 1+Y & 1-Y\\
 1-Y & 1+Y\\
  \end{array} \right)
\left(
\begin{array}{cc}
a_R \\
a^{\dag}_R\\
 \end{array} \right),
\end{equation}
where $X$ and $Y$ are coefficients related to the dynamical phases, $X=\frac{\cos(\phi_u)-\cos(2\phi_u)\cos(\phi_l)}{1-\cos(\phi_u)\cos(\phi_l)}$
and $Y=\frac{\sin(\phi_u)\sin(\phi_l)}{1-\cos(\phi_u)\cos(\phi_l)}$.
Finally, the transmission and Andreev reflection coefficients of Device B are analytically obtained,
\begin{eqnarray}
  \displaystyle T&=&T^\mathrm{CAR}=(1/4)X^2,\label{B_transT} \\
  \displaystyle R&=&(1/4)(1+Y)^2, \label{B_transR} \\
  \displaystyle T^\mathrm{LAR}&=&(1/4)(1-Y)^2.\label{B_transLAR}
\end{eqnarray}
Here the normal transmission coefficient $T$ is always the same as
the CAR coefficient $T^\mathrm{CAR}$,
which is well consistent with numerical results in Figs.7(b and c),
because the outgoing electron $b_L$ and the hole $b^\dag_L$ share
the same Majorana mode $\gamma_L$.
However, the reflective mode, $b_R$, is a mixture of two Majorana modes,
$\gamma_1$ and $\gamma_R$ [see Fig.1(b)].
Thus, it is no wonder that the probability of an outgoing electron differs from
that of a hole in Lead-2,
as is why the normal reflection coefficients $R$ is usually not equal
to the LAR coefficients $T^\mathrm{LAR}$ plotted in Figs.7(b and c).
From Eq.(\ref{Eq:LBformula}), this difference will eventually give rise
to an oscillating conductance to be shown later.
Furthermore, since there is only a Majorana fermion $\gamma_L$ outgoing to Lead-1,
the normal transmission and CAR coefficients are always less than $1/4$,
but the normal reflection and LAR coefficients can exceed $1/4$
due to two Majorana fermions $\gamma_R$ and $\gamma_1$ outgoing to Lead-2.

By setting $\phi_u=\phi+\frac{\pi}{2}$ and $\phi_l=\frac{\pi}{2}$,
the analytical transmission coefficients are plotted in Fig.7(d).
Here, analytical results accommodate well with numerical calculations in Fig.7(c).
Definitely, trajectories of Majorana fermions are
regulated by changing the dynamical phase $\phi$
which can be tuned by the upper or lower gate of the central QAHI.
Let us discuss the control of the propagating route of Majorana fermion $\gamma_2$
that always passes through the right TSC into the central QAHI.
When $\phi=\pi$, $R=1$ and $T=T^\mathrm{CAR}=T^\mathrm{LAR}=0$, namely, the Majorana fermion $\gamma_2$ is reflected at the left TSC-central QAHI interface with
$\gamma_L=0$ and $\gamma_R=\gamma_2$.
But at $\phi=\pi/2$,
$R=T=T^\mathrm{CAR}=T^\mathrm{LAR}=1/4$, namely, $\gamma_2$
totally ejects through the left TSC into Lead-1 with
$\gamma_L=\gamma_2$ and $\gamma_R=0$.
When $\phi=0$, $T^\mathrm{LAR}=1$, while $T=T^\mathrm{CAR}=R=0$, that is,
$\gamma_2$ gets a phase $\pi$ and is totally reflected with
$\gamma_L=0$ and $\gamma_R=e^{i\pi}\gamma_2$.
With other values of phase $\phi$, the incoming Majorana fermion $\gamma_2$
is controlled to leave partially into Lead-1 and to be reflected back partially.
Thus, by tuning the gate voltage, we do control the trajectory of chiral Majorana fermions.

\begin{figure}
  \centering
  \includegraphics[width=1\linewidth]{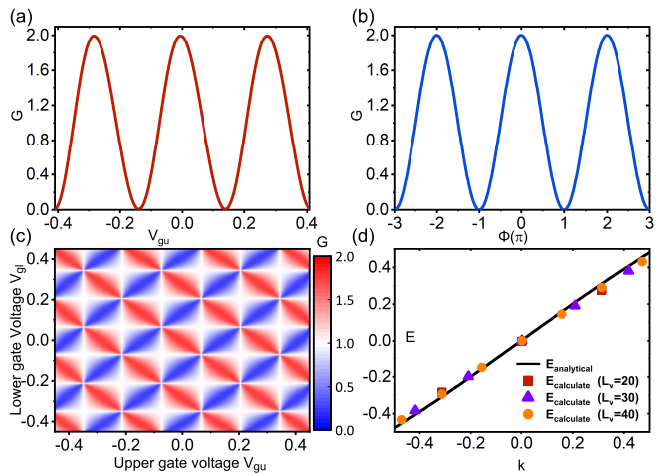}
  \caption{(color online)
  (a) and (b) are the linear conductance $G$ of Device B as a function of
  the upper gate voltage $V_{gu}$ and the dynamical phase $\phi$ from
  the numerical calculations and analytical results, respectively.
  (c) Colormap of conductance $G$ as a function of both
  the upper gate voltage $V_{gu}$ and lower gate voltage $V_{gl}$.
  (d) The numerical calculated energy dispersions from the oscillating conductance.
  Differently colored markers correspond to different length $L_v$ of the gate region.
  The black solid lines are the analytical energy dispersion relation from Fig.6(a).
  All the unmentioned parameters are the same as Figs.7(c and d).}
  \label{fig:fig8}
\end{figure}

Then we calculate the linear conductance $G$ of Device B,
which is derived from Eq.(\ref{Eq:LBformula}) as,
\begin{equation}
  G=\frac{dI}{dU}=\frac{e^2}{h}(T+T^\mathrm{CAR}+2 T^\mathrm{LAR}),
\end{equation}
where $U=V_2-V_1$ is the bias voltage between Lead-2 and Lead-1.
Combined with the relation in Eq.(\ref{ep.R}),
the linear conductance can be written in a plain form,
$G=\frac{e^2}{h}(1+T^\mathrm{LAR}-R)$.
Fig.8(a) shows the conductance $G$ versus the upper gate voltage $V_{gu}$, where $G$ periodically oscillates with the gate voltage $V_{gu}$
with the oscillation amplitude being $\frac{2e^2}{h}$.
In fact, although the normal transmission coefficient $T$
is always accordant with the CAR coefficient $T^\mathrm{CAR}$,
the normal reflection coefficient $R$ and the LAR coefficient $T^\mathrm{LAR}$
usually separate from each other [see Figs.7(b and c)],
so the conductance $G$ can take on a significant oscillation in Fig.8(a).

By using Eqs.(\ref{B_transT}-\ref{B_transLAR}),
the analytical expression of $G$ is explicitly denoted as
\begin{equation}
  G=\frac{e^2}{h}\left(1+\frac{\sin(\phi_u)\sin(\phi_l)}{1-\cos(\phi_u)\cos(\phi_l)}\right)
\end{equation}
When setting $\phi_u=\phi +\frac{\pi}{2}$ and $\phi_l=\frac{\pi}{2}$,
we plot the analytical conductance $G$ corresponding to
the dynamical phase $\phi$ in Fig.8(b), which is completely
consistent with the numerical curve in Fig.8(a).

Fig.8(c) shows the colormap of conductance $G$ as a function of
upper and lower gate voltages, $V_{gu}$ and $V_{gl}$.
The conductance periodically oscillates with both $V_{gu}$ and $V_{gl}$,
and it presents the equivalent effects of tuning the upper and the lower gate voltages.
Also, the conductance undulation appears a shuttle-like pattern in each period.
As shown by now, the oscillating conductance is experimentally observable.
Meanwhile, this quantum oscillation phenomenon can confirm the existence of chiral Majorana modes.
In addition, the oscillation phenomenon originates from
the dynamical phase $\phi$ controlled by gate voltages,
and can not be explained by any classical interpretation.
Given the half-integer quantum conductance plateaus
observed in the QAHI-TSC-QAHI system in experiments,\cite{Heqinglin2017science}
here are classical interpretations based on the percolation model
in two recent works.\cite{JiWenjieWenxiaogang2018PRL,HuangYingyi2018PRB,Qijunjie2019}
They consider that the superconductor in the QAHI-superconductor-QAHI system
is at the $\mathcal{N}=0$ and $\mathcal{N}=2$ phases,
but the $\mathcal{N}=1$ superconducting phase
(i.e. the TSC phase with the chiral Majorana edge modes) does not form.
Near the percolation threshold,
the incoming edge modes from the right QAHI could partially
be transmitted to the left QAHI through the superconductor region
with suitable leakage to adjacent chiral edges,
giving rise to a nearly flat half-integer conductance plateau.
In the present Device B, if based on the classical percolation model,
the QAHI-TSC-QAHI-TSC-QAHI junction can be equivalent to the coupling
of two QAHI-TSC-QAHI junctions.
Considering the mixing of $\mathcal{N}=0$ and $\mathcal{N}=2$ superconducting phases instead of the $\mathcal{N}=1$ TSC,
implementing a variable gate voltage along the edge of central QAHI regions
would have no influence on the transport through the QAHI-TSC-QAHI junctions
which eventually express a constant conductance.
So the oscillating behavior of conductance does serve
as smoking-gun evidence of the existence of the chiral Majorana fermions.

As with the same operation of transmission coefficients in Sec \ref{sec3},
the $E$-$k$ dispersion relation of chiral Dirac edge states of QAHI
could also be read out from the oscillation of the conductance versus the gate voltage.
But the oscillation period of the conductance is $2\pi$ as shown in Fig.8(b),
rather than $\pi$ in the transmission coefficient curves of Device A,
so the momentum $k_n =\frac{2 n \pi}{L_v}$ with the integer $n$.
Based on the numerical conductance in Fig.8(a),
the discrete data points $(E_n,\ k_n)$ can be obtained,
shown by red solid squares in Fig.8(d).
Also, the purple triangular and orange circle solid markers denote
the numerical $E$-$k$ relations corresponds to other gating
lengths $L_v =30a$ and $40a$, respectively.
The black dash line in Fig.8(d) represents the analytical $E$-$k$ relation from Fig.6(a).
Analytical and calculated results match well with each other,
no matter how long the gating region is.
So from the conductance oscillation, we can experimentally measure
the dispersion relation of the chiral Dirac edge state of the QAHI.
This method is effective regardless of the systematic parameters,
e.g. the length of the gating region and the chemical potentials $\mu_{\mathrm{tsc}}$ and $\mu_\mathrm{QAHI}$.

\section{\label{sec5}Results of Device C }

In this section, we propose an alternative design to manipulate
chiral Majorana edge modes with the conductance oscillation
to be observed in experiments, as Device C in Fig.1(c).
In comparison with Device A, weak coupling is introduced
between the central QAHI region and Lead-2.
The coupling strength can be tuned by the contact size
in order to mix the incoming Dirac edge modes $a_u$ with $a_d$
and redistribute the outgoing Dirac edge modes $b_R$ and $b_3$ [see Fig.1(c)].
Although Lead-3 is schematically diminished as a black square,
it is wide enough ($W_3=60a$) in calculations to avoid the direct mixing
between the incident and outgoing Dirac edge modes.
In addition, the results are independent of the specific position of
Lead-3, as long as it does not locate too close to Lead-2.
Compared with Device B, there are no more cyclic trajectories of
chiral Majorana fermions in Device C and the propagating route is clearer,
however, Device B should be easier to implement in experiments.

 \begin{figure}
\centering
\includegraphics[width=1.03\linewidth]{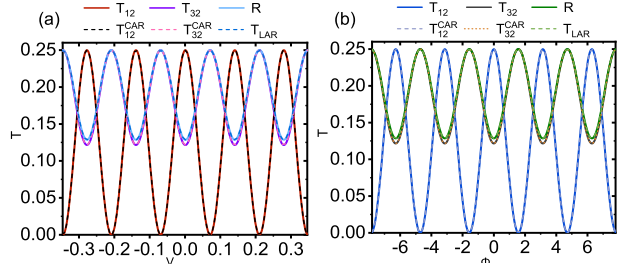}
\caption{(color online)
The transport properties of Device C.
(a) The normal transmission coefficients $T_{12}$ and $T_{32}$,
the CAR coefficients $T^\mathrm{CAR}_{12}$ and $T^\mathrm{CAR}_{32}$,
the normal reflection coefficient $R$, and
the LAR coefficient $T^\mathrm{LAR}$ as functions of the gate voltage $V_g$,
with the parameters $\mu_\mathrm{QAHI}=0.2$, $\mu_{\mathrm{tsc}}=0.55$ and $L_v=40a$.
(b) The analytical transmission coefficients from Eqs.(\ref{Eq:IIItrans1}-\ref{Eq:IIItrans33})
versus the dynamical phase $\phi$ with $\theta=1.07$ and an initial phase $\phi_0=3.26$
at the zero gate voltage.}
\label{fig:fig9}
\end{figure}

Now let us investigate the control of Majorana fermions
via Device C when the TSC is in the $\mathcal{N}=1$ phase.
Considering that a Dirac electron injects from Lead-2,
which is tantamount to two Majorana fermions $\gamma_1$ and $\gamma_2$.
One of the two Majorana fermions, $\gamma_2$, passes through the right TSC,
gets a dynamical phase $\phi$ in the gating QAHI region,
and splits two Majorana fermions $\gamma_L$ and $\gamma_3$ with
$\gamma_L=\gamma_2\sin\phi$ and $\gamma_3=\gamma_2\cos\phi$. The dynamical phase $\phi$ can control trajectories of Majorana fermion $\gamma_2$.
Then $\gamma_L$ passes through the left TSC into Lead-1
and $\gamma_3$ counterclockwise travels along the edge of central QAHI to a contact junction.
At the contact junction, the mode, $a_d=\frac{\sqrt{2}}{2}\gamma_3$, can partially
exit leftward into Lead-2 or rightward into Lead-3 [see Fig.1(c)].
Recall that the other Majorana fermion $\gamma_1$ is reflected at the boundary of QAHI-right TSC,
directly goes to the contact junction with $a_u=\frac{\sqrt{2}}{2}\gamma_1$,
and splits into two branches into Lead-2 and Lead-3.
Considering the current conservation,
the unitary scattering matrix at the contact junction can be set as
\begin{equation}
\left(\begin{array}{cc}b_R \\b_3\\ \end{array} \right)
= \left( \begin{array}{cc}
t & re^{-i\theta}  \\
-re^{i\theta}& t \\
  \end{array} \right)
\left(\begin{array}{cc}a_d \\a_u\\ \end{array} \right),
\end{equation}
where $t$ and $r$ respectively represent the tunneling amplitude and reflection amplitude
with $r^2+t^2=1$ and $\theta$ depicts an additional phase passing through the junction.
To be summarized, we list all six transmission coefficients of Device C as follows:

\begin{eqnarray}
  \displaystyle T_{12}&=&T^\mathrm{CAR}_{12}=(1/4)\sin^2\phi,\label{Eq:IIItrans1}\\
  \displaystyle T_{32}&=&[(t-r\cos\phi\sin\theta)^2+(r\cos\phi\cos\theta)^2]/4,\label{Eq:IIItrans2} \\
  \displaystyle T^\mathrm{CAR}_{32}&=&[(t+r\cos\phi\sin\theta)^2+(r\cos\phi\cos\theta)^2]/4,\label{Eq:IIItrans22}\\
  \displaystyle R_{2}&=&[(t\cos\phi+r\sin\theta)^2+(r\cos\theta)^2]/4,\label{Eq:IIItrans3} \\
  \displaystyle T^\mathrm{LAR}_{2}&=&[(t\cos\phi-r\sin\theta)^2+(r\cos\theta)^2]/4.\label{Eq:IIItrans33}
\end{eqnarray}
In the zero tunneling limit ($t=0$ and $r=1$),
the Majorana fermion $\gamma_3$ totally enters into Lead-3
while the mode $\gamma_1$ is totally reflected to Lead-2,
that is, Device C is reverted into Device A.
In the entire tunneling limit ($t=1$ and $r=0$), conversely,
$\gamma_3$ goes into Lead-2 while $\gamma_1$ goes into Lead-3.
Except for these two limiting cases, the four-side junction could mix $\gamma_3$ and $\gamma_1$,
leading to that $T_{32}\not=  T^\mathrm{CAR}_{32}$ and $R_{2} \not= T^\mathrm{LAR}_{2}$ as usual.

The numerical and analytical transmission and Andreev reflection coefficients
of Device C with moderate coupling are plotted in Figs.9(a and b).
For clear visibility, the tunneling probability $t^2$ and reflection strength $r^2$
of the contact junction are adjusted slightly away from 0.5.
Apparently, the normal transmission coefficients, normal reflection coefficient,
LAR and CAR coefficients, all periodically oscillate with the gate voltage $V_g$.
The oscillation amplitudes of these coefficients are quite large as shown in Fig.9(a).
It proves that the propagating route of the chiral Majorana fermion can well be
regulated and controlled by the electric gate.
In the analytical results in Fig.9(b),
we set $\theta=1.07$ and introduce an initial phase $\phi_0$ without gating, that is,
$\phi$ in Eqs.(\ref{Eq:IIItrans1}-\ref{Eq:IIItrans33})
is replaced by $\phi+\phi_0$ with $\phi_0=3.26$.
The analytical results are in perfect agreement with the numerical ones, see Figs.9(a and b), indicating that the transport of Majorana (Dirac) fermions along the chiral
edge state in TSC (QAHI) can well describe the transport process of the QAHI-TSC system.

Adopting the Landauer-B\"{u}ttiker formula in Eq.(\ref{Eq:LBformula})
and the same external circuit layout in Fig.1(a)
with the boundary conditions $I_1+I_2+I_3=0$ and $V_1=V_3$,
we can achieve the linear conductance $G_n$ of Device C.
Here the linear conductances $G_n$ is defined as
the ratio of the current $I_n$ to the bias $U=V_2-V_1$ between Lead-2 and Lead-1 (Lead-3).
Fig.10(a) displays the conductances $-G_1$, $G_2$ and $-G_3$
with the same parameter setting as Fig.9(a).
Clearly, they periodically oscillate as the gate voltage $V_g$ varies,
showing the manipulation of incoming Majorana fermions by tuning $V_g$.
The oscillating periods of the conductances are almost
identical and constant under the same $L_v$.
From the periodic oscillation of the conductance,
we can deduce the energy dispersion relation of the central QAHI marked
by red squares in Fig.10(b),
which is in good fit with the analytical dispersion relation.

Here we mention that conductance oscillation
could be experimentally observed,
which can confirm the existence of chiral Majorana edge modes
and show the effective control of their propagating route.
Devices B and C satisfy the aspiration of an oscillating observable.
Basically, ways of observing the oscillating conductance are identical in principle,
that is, to create a mixing between the incident two chiral Majorana fermions,
$\gamma_1$ and $\gamma_2$.
The mixing is realized by the cyclic propagating route of chiral Majorana edge modes
in the central QAHI region in Device B, while in Device C, 
a contact junction introduces the mixing.

\begin{figure}
  \centering
  \includegraphics[width=1\linewidth]{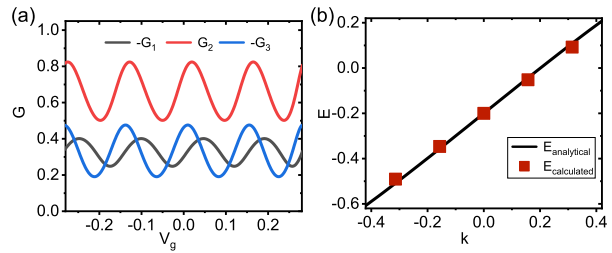}
  \caption{(color online)
(a) The conductances $G_n$ of Device C versus the gate voltage $V_g$
with the same external circuit layout in the inset in Fig.1(a).
The parameters are the same as Fig.9(a).
(b) The energy dispersion relation of QAHI from the oscillating conductance $G_n$.
The red squares and the black solid line are numerical and analytical results respectively.}
  \label{fig:fig10}
\end{figure}

\section{\label{sec6} the effect of potential disorder, the superconducting gap fluctuation
and superconducting phase difference on the conductance oscillations}

In the above, we have shown that the propagating route of chiral Majorana fermions
can well be controlled by the gate voltages,
which results in the oscillations of the conductance with the change of gate voltage.
In the real system, impurities and disorders exist inevitably,
which disrupt the transmissions of Majorana fermions and Dirac fermions.
In this section, we study the effect of the potential disorder,
the random spatial variation of the superconducting gap,
and the non-zero superconducting phase difference between two TSCs
on the oscillations of the conductance.

First, let us study the potential disorder.
In order to investigate the effect of the potential disorder,
we add an Anderson disorder term to the lattice Hamiltonian in Eq.(\ref{HQL}),
\begin{eqnarray}\label{HQL2}
  \mathcal{H}_{\mathrm{QAHI}} &=& \sum_\mathbf{i}\left[\psi_\mathbf{i}^{\dag} T_0 \psi_\mathbf{i}+
  \psi_\mathbf{i}^{\dag} w_\mathbf{i} \sigma_0 \psi_\mathbf{i}
  \right. \nonumber \\
   & &\left. +
  (\psi_\mathbf{i}^{\dag}T_x\psi_{\mathbf{i}+\delta \mathbf{x}}+\psi_\mathbf{i}^{\dag} T_y \psi_{\mathbf{i}+\delta \mathbf{y}})+\mathrm{H.c.}\right],
\end{eqnarray}
where the new term $\psi_\mathbf{i}^{\dag} w_\mathbf{i} \sigma_0 \psi_\mathbf{i}$ describes
the potential disorder.
At each site $\mathbf{i}$, $w_\mathbf{i}$ is uniformly distributed
in the interval [$-W_{\mathrm{sd}}$/2, $W_{\mathrm{sd}}$/2],
where $W_{\mathrm{sd}}$ denotes the strength of disorder.
Here we consider that the disorder exists in the whole central scattering.
Take Device B for example, the disorder can exist in the left TSC, the center QAHI and the right TSC regions.
For each non-zero $W_{\mathrm{sd}}$, the conductance curves are
averaged over 1000 random disorder configurations.
In fact, the average over the disorder configurations
is equivalent to the dephasing effect.\cite{Chenjiangchai2011JPCM}
In Sec.\ref{sec4}, we have shown that
the conductance of Device B oscillates with the gate voltage $V_{gu}$
in the absence of the disorder, see Fig.8(a).
Now we study the effect of disorder on the conductance oscillations.
Fig.11(a) shows the conductance $G$ of Device B versus the gate voltage $V_{gu}$
for the different disorder strength $W_{\mathrm{sd}}$.
In the weak disorder, the oscillations of the conductance,
including the oscillation amplitude and period, are almost not changed.
The conductance curve of the disorder strength $W_{\mathrm{sd}}=0.1$ (the yellow dash line in Fig.11(a))
almost coincides with the curve of $W_{\mathrm{sd}}=0$ (the red solid line in Fig.11(a)).
With the increase of the disorder strength $W_{\mathrm{sd}}$,
the amplitude of conductance oscillations slightly reduces.
But the conductance oscillations can well survive even
if the disorder strength $W_{\mathrm{sd}}$ reaches $0.4$,
which is close to the superconducting gap $\Delta$
and the mass gap $|m|$ of the QAHI.
In fact, the conductance oscillation originates from both
the chiral Dirac edge states in the QAHI region and the chiral Majorana edge modes in the TSC region,
so it is very robust against the disorder.
In particular, the period of the conductance oscillations
is almost not affected by the disorder. For example, the oscillation period
at $W_{\mathrm{sd}}=0.4$ is still equal to that at $W_{\mathrm{sd}}=0$.
So even in the strong disorder,
we can still obtain the dispersion relation of chiral Dirac
edge states of the QAHI from the oscillation curves of the conductance versus the gate voltage.

Next, we study the effect of the superconducting gap fluctuation
on the conductance oscillations.
In the experiments, it is difficult
to keep that the superconducting gap is perfectly uniform in real space.
Usually, there are some random fluctuations.
Here we consider that the superconducting gap in Eq.(\ref{HBDG}) depends
on the spatial site index $\mathbf{i}$.
The superconducting gap $\Delta_\mathbf{i}$ at the site $\mathbf{i}$ is assumed
to be uniformly distributed in the interval [$\Delta -\Delta_d/2$, $\Delta +\Delta_d/2$ ],
where $\Delta$ is the homogeneous superconducting gap without spatial variation and
$\Delta_{d}$ denotes the strength of spatial variation.
Fig.11(b) shows the conductance of Device B
versus the gate voltage $V_{gu}$ for each different $\Delta_{d}$.
One can clearly see that the conductance oscillation is robust against
the fluctuation of the superconducting gap in real space.
With the increase of the variation strength $\Delta_{d}$, the oscillation amplitude
slightly decreases and the oscillation period can almost be the same as that at $\Delta_{d}=0$.
When $\Delta_d=0.7$ (i.e. $\Delta_\mathbf{i}$ is uniformly distributed from $0$ to $0.7$),
the conductance oscillation can still survive.
If $\Delta_d$ increases further or the region with $\Delta_\mathbf{i}=0$ enlarges further,
the large region in the $\mathcal{N}=1$ TSC phase is tremendously destroyed, and then the conductance oscillation disappears.
These results indicate that the proposed scheme to manipulate the chiral Majorana fermions
is still effective when the superconducting gap is not uniform in real space.

\begin{figure}
    \centering
    \includegraphics[width=1\linewidth]{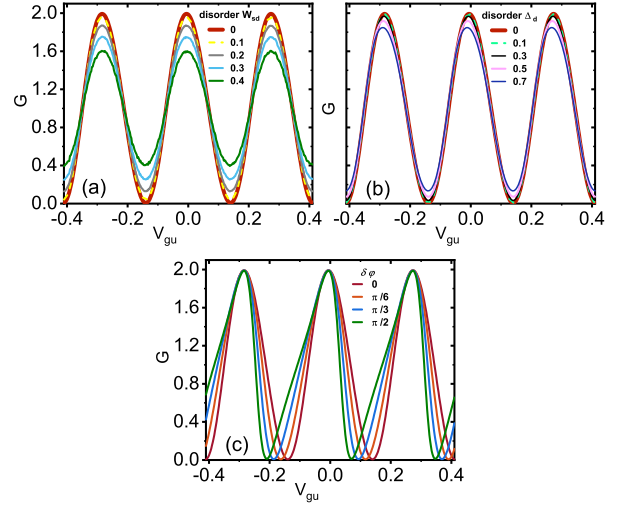}
\caption{
(a) and (b) are the linear conductance $G$ of Device B
versus the upper gate voltage $V_{gu}$ for the different disorder strength
$W_{\mathrm{sd}}$ and the different spatial variation of the superconducting gap $\Delta_{d}$, respectively.
For each $W_{\mathrm{sd}}$ or $\Delta_{d}$, the conductance curves are
averaged over 1000 random configurations.
(c) is $G$ of Device B versus $V_{gu}$ for
the different superconducting phase difference $\delta \varphi$.
The red solid curves in (a-c) are the same as the curve in Fig.8(a).
All other unmentioned parameters are the same as Fig.8(a).
    }
\end{figure}

In the above, we set that the superconducting phase
difference $\delta \varphi$ between two TSCs is zero for a clear description.
Below, let us study the effect of the non-zero $\delta \varphi$ on
the conductance oscillations.
When $\delta \varphi$ is non-zero, the superconducting gaps $\Delta$ in
the left and right TSC regions become $\Delta e^{i\varphi_L}$ and $\Delta e^{i\varphi_R}$,
and $\delta \varphi =\varphi_L -\varphi_R$.
In Fig.11(c), we show the linear conductance $G$ of Device B
as a function of the upper gate voltage $V_{gu}$ for each different $\delta \varphi$.
Except for the superconducting phase difference $\delta \varphi$,
the other parameters here are exactly the same as Fig.8(a).
One can see that the oscillation of the conductance versus the gate voltage
always exists regardless of the phase difference $\delta \varphi$.
It clearly indicates that even in the non-zero $\delta \varphi$,
our proposed scheme to control the propagating trajectories of Majorana fermions
is still effective.

Finally, we discuss the parameters in real materials.
In 2013, Chang {\sl et al.} have successfully realized the QAHI
in the Cr-$\mathrm{(Bi,Sb)_2Te_3}$ films.
In this material, the Fermi velocity $\hbar v_F $ is about
$260\ \mathrm{meV\ nm}$.\cite{Changcuizu2013science,WangjingZSC2015prb}
So we can determine the value of $A$, $A = \hbar v_F =260\ \mathrm{meV\ nm} $.
Then we choose the lattice constant as $a=0.26\ \mu \mathrm{m}$.
Under this condition, the dimensionless superconducting gap $\Delta=0.35$
corresponds to a real value $\Delta=0.35\ \mathrm{meV}$.
It is achievable because the proximity induced superconducting gap
of $\mathrm{Bi}_2\mathrm{Se}_3$ films on the $\mathrm{Nb}\mathrm{Se}_2$
substrate can reach $\Delta=0.5\ \mathrm{meV}$ even at 4.2K.\cite{Wangmeixiao2012Science,WangjingZSC2015prb}
Besides, the dimensionless mass gap $m=-0.5$ corresponds to a real value $m=-0.5 \mathrm{meV}$.
The QAH effect in Cr-$\mathrm{(Bi,Sb)_2Te_3}$ is measured at $\mathcal{T}=30\mathrm{mK}$,\cite{Changcuizu2013science}
the bulk gap of QAHI is of the order of meV, in the same magnitude as our approximation.
In addition, the bulk gap of QAHI depends on the thickness of the films, so $m$ can well be
tuned in experiments.
Also, the size of each TSC region (100$a$, 80$a$) corresponds to (26$\mu$m, 20.8$\mu$m),
which could be experimentally fabricated, compared to recent work
by He {\sl et al.}.\cite{Heqinglin2017science}
Besides, the size of gating QAHI region (20$a$, 20$a$) corresponds to (5.2$\mu$m, 5.2$\mu$m)
and the oscillating period of conductance is about 0.2 mV of Device C and 0.4 mV of Device B,
that is, the oscillation is visible when the scanning voltage is in the accuracy of $\mu$V.
In the above calculation, the temperature is set to zero. In this case,
the energy of the incident electron is fixed at $E=0$.
At a finite temperature $\mathcal{T}$, the energy of the incident electron
is distributed about in the range (-$k_B \mathcal{T}$, $k_B \mathcal{T}$).
Then the oscillation amplitude reduces because of
the participation of non-zero energy incident electrons,
but the oscillation period can still remain the same as that at the zero temperature.
At the low finite temperature
(e.g. the temperature is an order of magnitude lower than the superconductor gap),
the conductance oscillation can be clearly visible.
So the oscillations of the conductance should be experimentally observed
in the present technologies.

\section{\label{sec7}conclusion}

In summary, we design three devices to manipulate chiral Majorana edge modes,
within the framework of QAHI-TSC-QAHI-TSC-QAHI junction.
The non-equivalent topology of the $\mathcal{N}=1$ TSC and
the $\mathcal{C}=1$ QAHI separates the two Majorana modes,
derived from the incoming regular electron mode.
Then via the external gate voltage, a dynamical phase is induced
to control trajectories of
the chiral Majorana fermion in the planned Device A,
leading to the oscillation of transmission coefficients.
Moreover, the following Devices B and C reach an achievement of observable
oscillation of the conductance, which could be detected in real experiments.
The oscillating conductance is robust against the disorder and
the random spatial variation of the superconducting gap
due to the topological protection of chiral edge states.
The stable oscillation period could also be utilized to deduce
the energy dispersion relation of the chiral Dirac edge mode of the QAHI region.
In addition, this periodically oscillating conductance versus the gate voltage
conspicuously verify the existence of chiral Majorana edge modes,
which would never appear in any classical interpretation.

\section*{ACKNOWLEDGMENTS}

This work was financially supported by National Key R and D Program of China (Grant No. 2017YFA0303301),
NSF-China (Grant Nos. 11574007 and 11921005),
the Strategic Priority Research Program of Chinese Academy of Sciences (Grant No. XDB28000000),
and Beijing Municipal Science \& Technology Commission No.Z181100004218001.

%

\end{document}